\documentclass[3p,,preprint,12pt]{elsarticle}
\makeatletter\if@twocolumn\PassOptionsToPackage{switch}{lineno}\else\fi\makeatother

\usepackage{tabulary,xcolor}
\usepackage{amsfonts,amsmath,amssymb}
\usepackage[T1]{fontenc}
\makeatletter
\let\save@ps@pprintTitle\ps@pprintTitle
\def\ps@pprintTitle{\save@ps@pprintTitle\gdef\@oddfoot{\footnotesize\itshape \null\hfill\today}}
\def\hlinewd#1{%
  \noalign{\ifnum0=`}\fi\hrule \@height #1%
  \futurelet\reserved@a\@xhline}

\AtBeginDocument{\ifNAT@numbers \biboptions{sort&compress}\fi}

\makeatother

\usepackage{ifluatex}
\ifluatex
\usepackage{fontspec}
\defaultfontfeatures{Ligatures=TeX}
\usepackage[]{unicode-math}
\unimathsetup{math-style=TeX}
\else 
\usepackage[utf8]{inputenc}
\fi 
\ifluatex\else\usepackage{stmaryrd}\fi

\usepackage{url,multirow,morefloats,floatflt,cancel,tfrupee}
\makeatletter

\AtBeginDocument{\@ifpackageloaded{textcomp}{}{\usepackage{textcomp}}}
\makeatother
\usepackage{colortbl}
\usepackage{xcolor}
\usepackage{pifont}
\usepackage[nointegrals]{wasysym}
\makeatletter

\def\mcWidth#1{\csname TY@F#1\endcsname+\tabcolsep}

\def\cAlignHack{\rightskip\@flushglue\leftskip\@flushglue\parindent\z@\parfillskip\z@skip}
\def\rAlignHack{\rightskip\z@skip\leftskip\@flushglue \parindent\z@\parfillskip\z@skip}

\@ifundefined{etal}{}{}

\usepackage{ifxetex}
\ifxetex\else\if@twocolumn\@ifpackageloaded{stfloats}{}{\usepackage{dblfloatfix}}\fi\fi

\AtBeginDocument{
\expandafter\ifx\csname eqalign\endcsname\relax
\def\eqalign#1{\null\vcenter{\def\\{\cr}\openup\jot\m@th
  \ialign{\strut$\displaystyle{##}$\hfil&$\displaystyle{{}##}$\hfil
      \crcr#1\crcr}}\,}
\fi
}

\AtBeginDocument{%
  \@ifpackageloaded{endfloat}%
   {\renewcommand\efloat@iwrite[1]{\immediate\expandafter\protected@write\csname efloat@post#1\endcsname{}}}{\newif\ifefloat@tables}%
}%

\def\BreakURLText#1{\@tfor\brk@tempa:=#1\do{\brk@tempa\hskip0pt}}
\let\lt=<
\let\gt=>
\def\processVert{\ifmmode|\else\textbar\fi}

\@ifundefined{subparagraph}{
\def\subparagraph{\@startsection{paragraph}{5}{2\parindent}{0ex plus 0.1ex minus 0.1ex}%
{0ex}{\normalfont\small\itshape}}%
}{}

\newcommand\role[1]{\unskip}
\newcommand\aucollab[1]{\unskip}
  
\@ifundefined{tsGraphicsScaleX}{\gdef\tsGraphicsScaleX{1}}{}
\@ifundefined{tsGraphicsScaleY}{\gdef\tsGraphicsScaleY{.9}}{}
\def\checkGraphicsWidth{\ifdim\Gin@nat@width>\linewidth
	\tsGraphicsScaleX\linewidth\else\Gin@nat@width\fi}

\def\checkGraphicsHeight{\ifdim\Gin@nat@height>.9\textheight
	\tsGraphicsScaleY\textheight\else\Gin@nat@height\fi}

\def\fixFloatSize#1{}
\let\ts@includegraphics\includegraphics

\def\inlinegraphic[#1]#2{{\edef\@tempa{#1}\edef\baseline@shift{\ifx\@tempa\@empty0\else#1\fi}\edef\tempZ{\the\numexpr(\numexpr(\baseline@shift*\f@size/100))}\protect\raisebox{\tempZ pt}{\ts@includegraphics{#2}}}}

\AtBeginDocument{\def\includegraphics{\@ifnextchar[{\ts@includegraphics}{\ts@includegraphics[width=\checkGraphicsWidth,height=\checkGraphicsHeight,keepaspectratio]}}}

\DeclareMathAlphabet{\mathpzc}{OT1}{pzc}{m}{it}

\def\URL#1#2{\@ifundefined{href}{#2}{\href{#1}{#2}}}

\def\UrlOrds{\do\*\do\-\do\~\do\'\do\"\do\-}%
\g@addto@macro{\UrlBreaks}{\UrlOrds}

\edef\fntEncoding{\f@encoding}

\makeatother

\newif\ifmultipleabstract\multipleabstractfalse%
    \makeatletter
\def\ead{\@ifnextchar[{\@uad}{\@ead}}
\gdef\@ead#1{\bgroup
   \def\_{\string\underscorechar\space}
   \def\{{\string\lbracechar\space}
   \def\textdagger{\string\textdagger\space}
   \def\texttildeapprox{\string\texttildeapprox\space}
   \def~{\hashchar\space}
   \def\}{\string\rbracechar\space}
   \edef\tmp{\the\@eadauthor}
   \immediate\write\@auxout{\string\emailauthor
     {#1}{\expandafter\strip@prefix\meaning\tmp}}
  \egroup
}
\gdef\emailauthor#1#2{\stepcounter{ead}
      \g@addto@macro\@elseads{\raggedright
      \let\corref\@gobble
      \eadsep\texttt{#1} (#2)
      \def\eadsep{\unskip,\space}}
}

\makeatother
  
\begin{document}

\begin{frontmatter}

\title{Probing multi-mobility edges in quasiperiodic mosaic lattices}
    
\author[1]{{Jun Gao}\corref{cor1}%
 \fnref{fn1}}
\ead{junga@kth.se}
\author[2,3]{{Ivan M. Khaymovich}\corref{cor1}%
\fnref{fn1}}
\ead{ivan.khaymovich@gmail.com}
\author[4]{Xiao-Wei Wang}
\author[1]{Ze-Sheng Xu}
\author[1]{Adrian Iovan}
\author[1]{Govind Krishna}
\author[1]{Jiayidaer Jieensi}
\author[1]{Andrea Cataldo}
\author[2,5]{Alexander V. Balatsky}
\author[1]{Val Zwiller}
\author[1]{{Ali W. Elshaari}\corref{cor1}}
\ead{elshaari@kth.se}

\affiliation[1]
{organization={Department of Applied Physics, KTH Royal Institute of Technology, Albanova University Centre}, city={Stockholm}, postcode={106 91}, country={Sweden}}
\affiliation[2]
{organization={Nordita, Stockholm University and KTH Royal Institute of Technology}, city={Stockholm}, postcode={SE-106 91}, country={Sweden}}
\affiliation[3]
{organization={Institute for Physics of Microstructures, Russian Academy of Sciences},  city={Nizhny Novgorod}, postcode={603950}, country={Russia}}
\affiliation[4]
{organization={Center for Integrated Quantum Information Technologies (IQIT), School of Physics and Astronomy and State Key Laboratory of Advanced Optical Communication Systems and Networks, Shanghai Jiao Tong University}, city={Shanghai}, postcode={200240}, country={China}}
\affiliation[5]
{organization={Department of Physics, University of Connecticut}, city={Storrs, Connecticut}, postcode={06269}, country={USA}}

\cortext[cor1]{Corresponding author}
\fntext[fn1]{These authors contributed equally to this work}

\begin{abstract}
 The mobility edge (ME) is a crucial concept in understanding localization physics, marking the critical transition between extended and localized states in the energy spectrum. Anderson localization scaling theory predicts the absence of ME in lower dimensional systems. Hence, the search for exact MEs, particularly for single particles in lower dimensions, has recently garnered significant interest in both theoretical and experimental studies, resulting in notable progress. However, several open questions remain, including the possibility of a single system exhibiting multiple MEs and the continual existence of extended states, even within the strong disorder domain. Here, we provide experimental evidence to address these questions by utilizing a quasiperiodic mosaic lattice with meticulously designed nanophotonic circuits. Our observations demonstrate the coexistence of both extended and localized states in lattices with broken duality symmetry and varying modulation periods. By single site injection and scanning the disorder level, we could approximately probe the ME of the modulated lattice. These results corroborate recent theoretical predictions, introduce a new avenue for investigating ME physics, and offer inspiration for further exploration of ME physics in the quantum regime using hybrid integrated photonic devices.
\end{abstract}
    
  \end{frontmatter}

\section{Introduction}

Disorder-induced localization, a phenomenon initially predicted by P. W. Anderson in 1958 \cite{Anderson1958}, has been a prominent topic in condensed matter physics \cite{disrev1,disrev2}. The scaling theory of localization \cite{scaling} revealed that in lower-dimensional disordered systems, all states become localized, whereas in three-dimensional systems, localized and extended eigenstates can coexist, resulting in the existence of a critical energy $E_c$ known as the mobility edge (ME) \cite{Evers2008}. Notably, when a quasiperiodic potential replaces random disorder, such as in the Aubry-Andr\'e (AA) model \cite{AAH1,AAH2}, a distinct picture emerges. This model suggests an energy-independent critical metal-insulator transition at a self-dual point, subsequently confirmed by experiments conducted both on photonic \cite{AAH3} and atomic \cite{AAH4,AAH5} platforms. However, due to its self-dual symmetry, the AA model does not possess a ME. The existence of MEs in one-dimensional (1D) systems is primarily conjectured in more generalized models \cite{Prange1983,Biddle2009,Biddle2010,Gopalakrishnan2017,Celardo2016,Deng2019,Saha2019,Deng2018,Nosov2019,Nosov2019-2,Kutlin2020,Deng2022,Das2022beta-ens,Das2023beta-ens_IMK,DasSarma1988,DasSarma1990,Ganeshan2015,Li2017,Yao2019,Yin2020,Roy2018,Danieli2015,Ahmed2022,Lee2022,Kim2022}, serving as a catalyst for experimental investigations based on ultracold atoms \cite{Ucold1,Ucold2,Ucold3}.

Recently, a significant advancement in the field of ME physics has emerged with the introduction of Avila's global theory \cite{Avila2015}, one of his {\it Fields Medal} work. This novel theoretical framework has uncovered a distinct class of exactly solvable 1D models, where quasiperiodic on-site potentials are incorporated with certain periods \cite{Wang2020}. Referred to as mosaic lattices, these models exhibit a range of compelling features. Notably, unlike previous models employing random or other quasiperiodic disorders, the mosaic lattice displays the remarkable property of hosting multiple MEs while breaking self-duality symmetry. Moreover, regardless of the strength of the quasiperiodic potential, extended states persist throughout the system –– a striking departure from the previous findings.

Here, we conducted experimental implementation of quasiperiodic mosaic lattices using integrated silicon nitride ($\text{Si}_3\text{N}_4$) photonic circuits with complementary metal-oxide-semiconductor (CMOS) compatible fabrication technology \cite{TPRev1,TPRev2,TPRev3,TPRev4,InteP,Elshaari2020,Moody2022,Chang2023}. By precisely engineering the on-site potential of each lattice site and adjusting the in-between gaps, while maintaining uniform hopping terms, we successfully achieved the desired quasiperiodic modulation over a wide tuning range at room temperature. Through single-site excitation of the photonic mosaic lattice in the strong tuning regime, we observe clear signatures of multiple MEs, which arise from the energy-dependent coexistence of both extended and localized states in the system. The existence of MEs is further confirmed by scanning the quasperiodic potential strength and probing the average energy of the injected state. Our results showcase the capacity of integrated photonics platforms to investigate ME physics in a scalable and precise manner, with the potential to explore and uncover unique quantum features depending on bosonic coalescence \cite{Lahini2010,Segev2013,Crespi2013} in quasiperiodic lattice models.

\section{Theoretical Results}

\begin{figure*}[!t]
	\centering
	\includegraphics[width=0.95\linewidth]{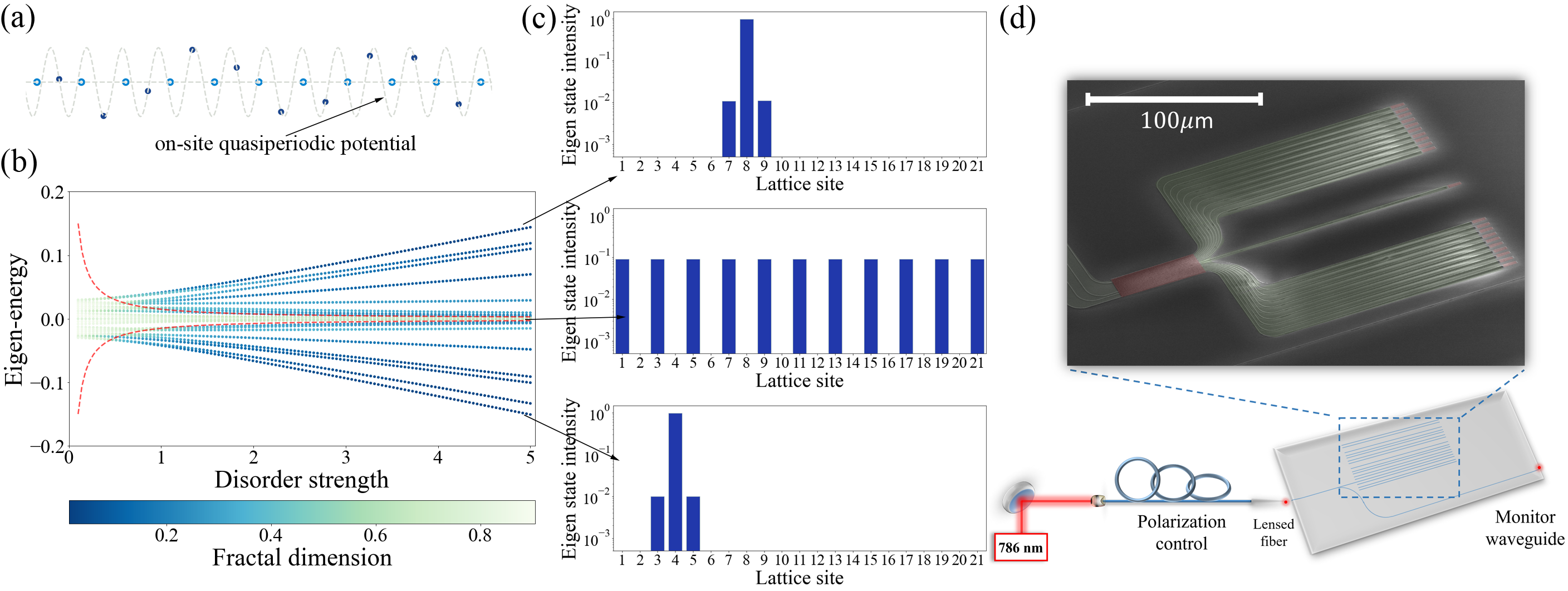}
	\caption{\textbf{Mobility edge in $\kappa=2$ quasiperiodic mosaic lattice.} 
		{(a)~Schematic of the 1D quasiperiodic mosaic lattice lattice, exemplifying the mobility edge phenomena in condensed matter physics. For the case of $\kappa=2$, the energy of every second lattice site is modified in accordance with equation \ref{eq:2}, while the hopping constant $J$ between lattice sites is held constant throughout. (b)~Energy diagram illustrating the dependence of eigenstate energies on quasiperiodic potential strength ($\lambda$). Red dashed lines show the location of the mobility edges. (c)~Real-space distribution of three distinct eigenstates at a disorder strength $\lambda/J=5$: (i) the highest positive energy eigenstate, localized predominantly at site 8; (ii) an extended state near zero energy; and (iii) the highest negative energy eigenstate, localized primarily at site 4. The system exhibits two mobility edges, as described by equation~\eqref{eq:5}, where eigenstates transition from localized to extended upon crossing the mobility edge. (d)~Sketch of the experimental setup and the SEM image of the fabricated device. The scale bar corresponds to 100 \textmu m.}
	}
	\label{fig1}
\end{figure*}

The Hamiltonian of the quasiperiodic mosaic model can be described as
\begin{equation}\label{eq:1}
H  =J \sum_j\left(c_j^{\dagger} c_{j+1}+\text { H.c. }\right)+2 \sum_j \lambda_j n_j. \\
\end{equation}
Here, $c_j^{\dagger}$ is the creation operator at site $j$, $J$ is the nearest neighbor hopping term, and $\lambda_j$ represents the on-site quasiperiodic potential modulation, which is given by the following formula, 
\begin{equation}\label{eq:2}
\lambda_j  = \begin{cases}\lambda \cos [2 \pi(\omega j+\theta)], & j=\kappa m, \\
0, & \text { otherwise }.\end{cases}
\end{equation}
$\theta$ is the phase offset during the modulation, $\omega$ is an irrational number, for instance $(\sqrt{5}-1)/2$ in our case, and $\kappa$ is an integer determining the mosaic modulation period. When $\kappa=1$, the lattice reduces to the AA model with self-dual symmetry and transition at $\lambda/J=1$, whereas when $\kappa\neq1$, the duality symmetry of the lattices is broken. $m$ refers to is an integer running from 1 to $N$, where $N$ is the quasi-cell number. Fig.~\ref{fig1}(a) demonstrates a 1D quasiperiodic mosaic model with $\kappa=2$ modulation period. In our simulation and experimental design, we set $\theta=0$ for the convenience, and $J=0.015$ \textmu m$^{-1}$ enabling wide tuning range of the potential amplitude modulation $\lambda_j$. 

Following Avila’s profound global theory \cite{Avila2015}, it has been theoretically proved \cite{Wang2020} that the mosaic model indeed manifests energy-dependent MEs by computing the Lyapunov exponent (see Supplemental Information (SI)~A for details), which can be described by the following expression:

\begin{eqnarray}\label{eq:3}
\left|\lambda a_\kappa\right|=J \quad\text{ for }\quad E&=&E_c \quad\text{, with} \\
\label{eq:4}
a_\kappa = \frac{\sin(\kappa p)}{\sin p}, \quad E &=& 2J \cos p 
\end{eqnarray}
the parametrization of the energy $E$ via the real (imaginary)-valued momentum $p$ for $|E|\leq 2 J$ ($|E|>2 J$).
A mosaic lattice with $\kappa$ modulation period hosts $2(\kappa-1)$ MEs, which are distributed in energy spectrum 
around $E=2J \cos(\pi m/\kappa)$, where $a_\kappa = 0$ and the extended states survive at arbitrarily strong potential $\lambda$, which is a new fundamental feature of mosaic lattices (see SI~A.1). For the simplest yet nontrivial case of $\kappa=2$, the two MEs are given by ~\cite{Wang2020}
\begin{equation}
E_c= \pm {J^2}/{\lambda}.\label{eq:5}
\end{equation}
In Fig.~\ref{fig1}(b) we show the calculated eigenvalues of a 21-site lattice versus the modulation strength $\lambda$ based on our chosen parameters. To characterize the MEs, we utilize the finite-size fractal dimension~\cite{Evers2008,Dq_Arnd}, defined as $D_2=-\ln (IPR) / \ln L$ via the inverse participation ratio $\mathrm{IPR}=\sum_{j} |\psi_{i}(j)|^4$ of eigenstates $\psi_{i}(j)$, corresponding to the specific energy $E_i$, to distinguish extended ($D_2 \rightarrow 1$) and localized states ($D_2 \rightarrow 0$) at large $L$. The dashed blue lines represent the exact MEs for $\kappa=2$ mosaic lattice as the transition between $D_2=0$ and $1$. Numerical simulations with larger sample sizes can be found in SI~B.

In Fig.~\ref{fig1}(c), we showcase three distinct eigenstates intensity distribution at a disorder level of $\lambda/J=5$. We could see the spatial distributions of the wave functions are exponentially localized at disordered sites $j=\kappa m$ (shown for the highest and lowest energy). For the extended eigenstate, the particle tends to stay at the sites without potential modulation (see SI~A.2), this also explains the survival of extended states in the mosaic lattice at strong potential.

\section{Experimental Implementation}

\begin{figure*}[!t]
	\centering
	\includegraphics[width=0.95\linewidth]{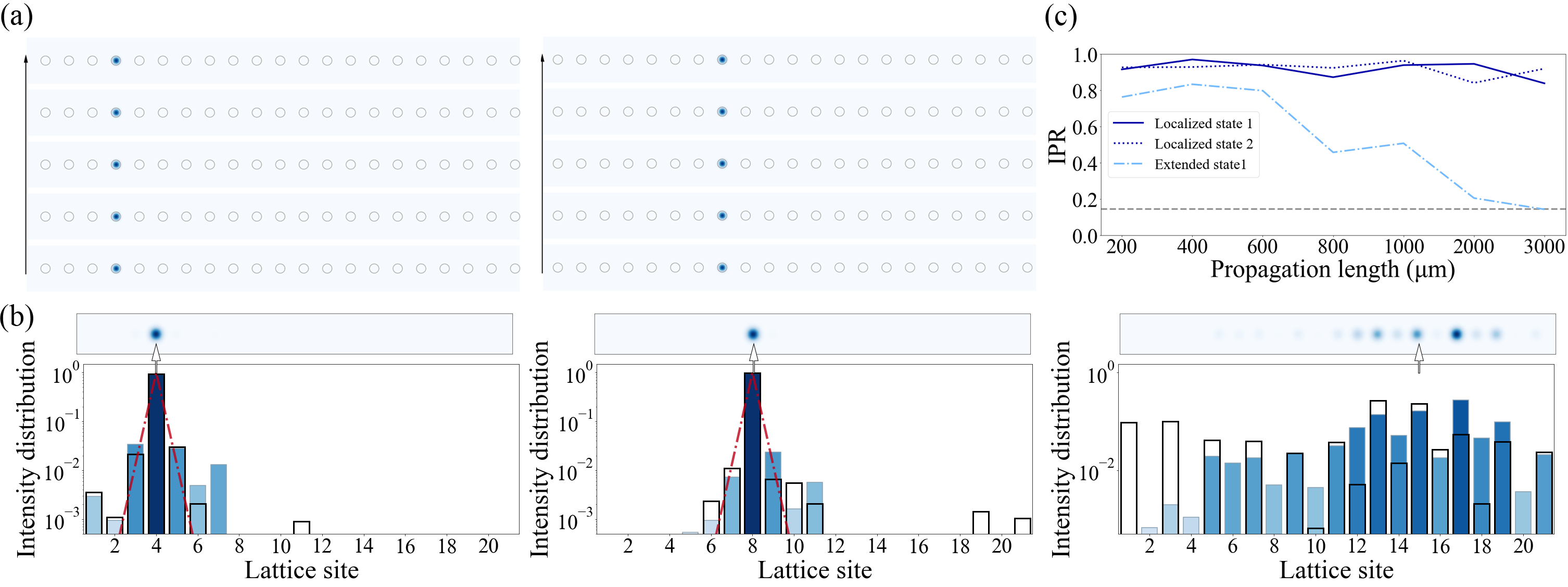}
	\caption{\textbf{Experimental probe of localized and extended states in $\kappa=2$ mosaic lattice.}  
	{(a)~Real-space distribution of light intensity probed every 200 \textmu m along the lattice, with light injected at lattice site 4. The observed confinement of light primarily at site 4 with minimal spreading to neighboring waveguides indicate the strong localization due to overlap with the localized lowest energy eigenstate depicted in Fig.\ref{fig1}(c). Similar measurement for light injected at waveguide 8, corresponding to the highest energy localized eigenstate depicted in Fig.\ref{fig1}(c). The observed localization of light, consistent with the behavior described in Fig.\ref{fig2}(a), emphasizes the strong overlap with the localized states. (b)~Real-space distribution of light intensity for light injected at site 4, 8 and 15, measured after 3000 \textmu m of light propagation in the lattice. The extended state exhibits significant wave-packet spreading. Bar plot below shows the light intensity on a logarithmic scale, unfilled bars represent simulation results and colored bars indicate experimental measurements, with the color code highlighting the light intensity. The dash-dotted red straight lines show the theoretically predicted exponential wave-function decay with the Lyapunov exponent $\gamma_0(E) = \ln|\lambda a_\kappa|/\kappa$, taken at the corresponding energy. The white arrows point out the input sites. (c)~Inverse participation ratio (IPR) calculated for the three wave packets, presented in Figs.\ref{fig2}(a), and (b). Single-site excitations, overlapping strongly with localized states, maintain a high level IPR for varying propagation lengths, while the one, overlapping with an extended state, as the right panel in Fig.\ref{fig2}(b), exhibits IPR reduction with the propagation distance, reaching 0.14 at longer propagation length. This behavior confirms the ME presence for a specific disorder strength $\lambda$.}
	}
	\label{fig2}
\end{figure*}

In our experiment, we implement the photonic quasiperiodic mosaic lattices based on integrated $\text{Si}_3\text{N}_4$ photonics platform (see SI~C for more fabrication details) \cite{silicon}. A scanning electron microscope (SEM) image of the nanophotonic device is presented in Fig.~\ref{fig1}(d). To design the desired on-site potential of each modulated site, we control the width of each waveguide according to numerical vectorial mode solver. We set 550~nm as the default width for a flexible tuning range while maintaining single-mode profile operation, and the modulated sites are designed to yield the potential modulation at a given modulation level of $\lambda/J=5$, shown in SI~A.1 to be enough to form all the MEs. The waveguide separation is carefully designed to keep the hopping term uniform due to the asymmetric coupling of different waveguide widths (see SI~D for the design methods and SI~F for device charaterization). We choose the 4th, 8th and 15th inputs to probe different regimes in the energy diagram. We adiabatically expand the output array by a fan-out structure, and all the output waveguides are coupled to grating couplers for the spatial intensity measurement. An additional monitor waveguide is fabricated for the facet beam profile imaging and polarization control as our previous designs~\cite{Gao2022,Xu2022,gao2023scalable}. We also vary the propagation lengths in different samples (with the same fabrication recipe) from 200 to 1000 \textmu m with an interval of 200 \textmu m to probe the light dynamics in the lattices.

The schematic of the experimental setup is shown in Fig.~\ref{fig1}(d). The photonic quasiperiodic mosaic lattice is probed using a coherent laser at a wavelength of 786~nm, which is prepared with horizontal polarization (TE mode). The light is coupled to the lattice through a lensed fiber mounted on a 6-axis nano-positioning stage (Thorlabs NanoMax). The input waveguide is divided into two paths: one serves as a monitor waveguide for polarization control, while the other leads to the injection site. The output intensity is top-imaged using a 40X objective and directly measured by a charge-coupled device (CCD) camera that records the reflected light from the grating couplers. An example of image acquisition can be found in the SI~E, Fig. S3. 

We first probe the two localized states in $\kappa=2$ photonic mosaic lattice. Figs.~\ref{fig2}(a) presents the top images of light intensity distribution of localized states every 200 \textmu m along the propagation distance. The grey circles mark the position of every site in the lattice. The figures clearly show the strong spatial confinement of the injected light at site 4 and site 8. These two excitation cases correspond to the lowest and highest energy eigenstates, shown in Fig.~\ref{fig1}(c). We also calculate the theoretical predictions of the exponential wave-function decay, based on the Lyapunov exponent calculation $\gamma_0 = \ln|\lambda a_\kappa| / \kappa$ to the wave-packet intensities in Fig.~2(b), $|\psi_i(j)|\sim e^{-\gamma_0 |j-j_i|}$, where $\max_j |\psi_i(j)| = |\psi_i(j_i)|$. A good agreement with the experimental data shows that the wave-packet dynamics at our propagation length gives already a reasonable approximation for the corresponding localized eigenstates. To probe an extended state in the lattice, we choose the 15th site as the excitation, based on the overlap of the eigenstates with the site-injected wave packets. In order to demonstrate a stark contrast between localized and extended states, we additionally prepared samples with even longer propagation lengths (2000 and 3000 \textmu m). Fig.~\ref{fig2}(b) illustrates the light intensity distribution in the lattice over a propagation distance of 3000 \textmu m with both linear (top) and logarithmic (bottom histogram) scale, confirming truly delocalized behavior of the extended state. To quantify the transport behavior, we calculate the IPR values for spreading wave-packets in all three cases versus propagation distance. The IPR values in Fig.~\ref{fig2}(c) for the localized cases remain close to 1 as the propagation length increases, while the IPR value for the extended state shows the expansion of the initial single-site wave packet (see SI~A.3) and the value decreases below 0.15 at long distance. More data can be found in SI~H to showcase the extended behavior in $\kappa=2$ photonic mosaic lattice. Our measurements provide experimental evidence of energy-dependent localization phase transition in the mosaic model.

\begin{figure*}
	\centering
	\includegraphics[width=0.95\linewidth]{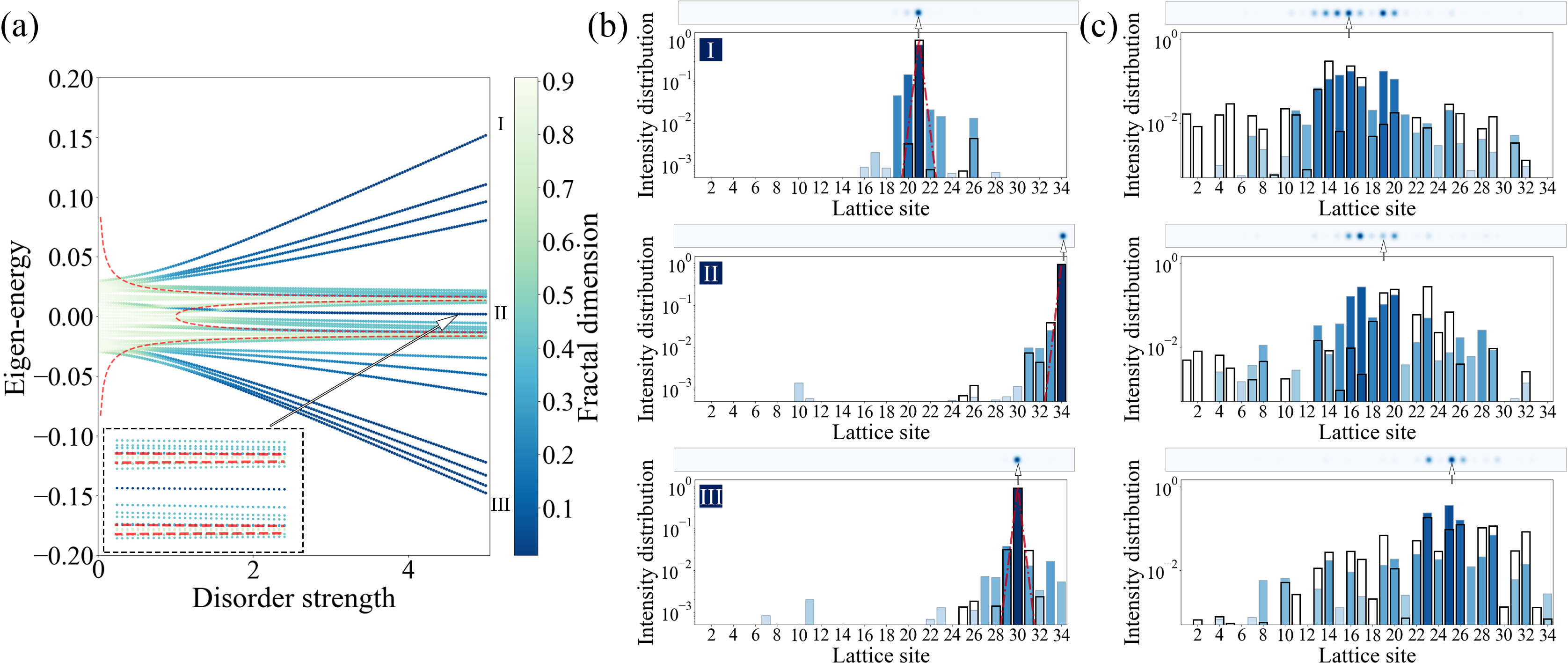}
	\caption{\textbf{Experimental probe of $\kappa=3$ mosaic lattice of 34 sites.} 
		{(a)~Eigenvalue evolution with increasing quasiperiodic disorder strength $\lambda$. Each point represents the energy of an individual eigenstate, color coded by the corresponding fractal dimension. As the disorder strength increases, the system develops four MEs (red dashed lines), separating three localization regimes \textbf{I}, \textbf{II}, \textbf{III}, see labeled panels, with two extended eigenstates regimes in between highlighted by a fractal dimension approaching unity. The inset shows the fine structure of 4 MEs under strong modulation strength. (b)~Propagation of a single-site excitation probes different regimes of the energy spectrum at a disorder strength of $\lambda/J=5$. The eigenstates distributions in real space and their overlap with the corresponding choice of single site excitation are given in the \textbf{SI~G}. \textbf{I, II, III} present the output spatial distribution of the light intensity in a logarithmic scale after propagating 3000 \textmu m in the lattice. Unfilled bars represent simulation results and colored bars indicate experimental measurements. The dash-dotted red straight lines show the theoretically predicted exponential wave-function decay with the Lyapunov exponent $\gamma_0(E) = \ln|\lambda a_\kappa|/\kappa$, taken at the corresponding energy. The experiment shows good agreement with both the analytical prediction and the simulation, where the light is exponentially localized in the excitation lattice site. The white arrows indicate the input sites. (c)~Experimental and numerical intensity distributions after propagating 3000 \textmu m in the lattice for a single-site excitation at the 16th, 19th and 25th lattice site respectively, overlapping with the extended eigenstates in the energy spectrum. More spreading of light highlights delocalization transition in the eigenstate spectrum, thus, exhibiting a ME. The data is given in both linear and logarithmic scale.}
    }
	\label{fig3}
\end{figure*}

\begin{figure}
	\centering
	\includegraphics[width=0.95\linewidth]{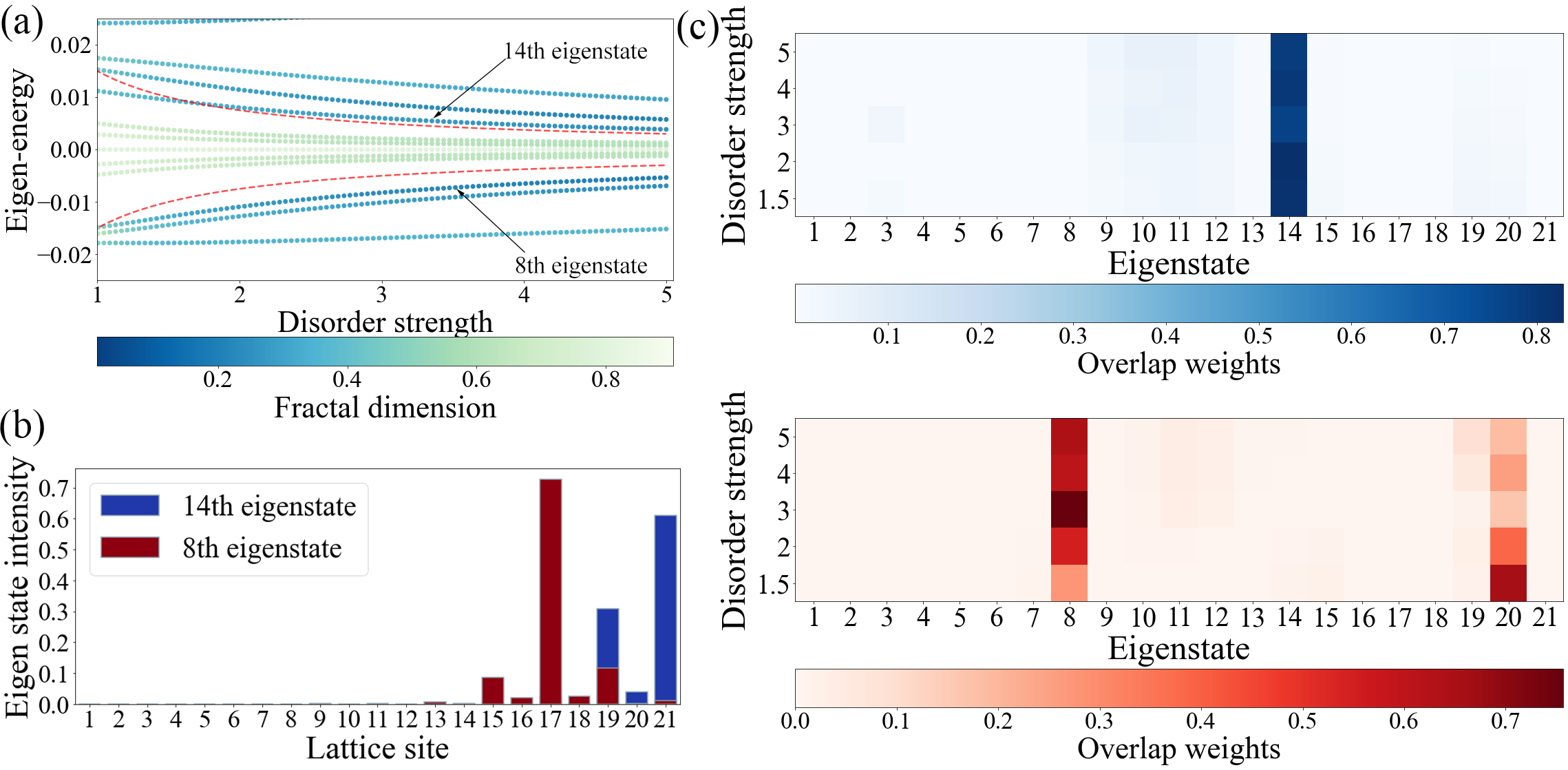}
	\caption{\textbf{Probing the critical behavior near mobility edges in $\kappa=2$ mosaic lattice.}
		{(a)~provides a detailed view of the critical 8th and 14th eigenstates relative location in the broader energy spectrum with respect to the MEs (red dashed lines). It is noted that these eigenstates reside in close proximity to the lower and upper branches of the exact ME, as determined by the theoretical model. The critical energy $E_c$ is given by $E_c = {J^2}/{\lambda}$, wherein the ME occurs. (b)~illustrates the spatial distribution of the 8th and 14th eigenstates that are found to have a significant overlap with the 17th and 21st lattice sites, respectively. These specific sites are chosen for the application of single-site excitation to effectively probe the system. (c)~The color maps present the weight overlap analysis between the eigenstates and the output intensity distribution, obtained experimentally following the single-site excitation at the 21st and 17th waveguide and after the wave has propagated a distance of 100 \textmu m through the lattice structure. It is observed that the output intensity distribution shows a considerable overlap with the 14th and 8th eigenstates, positioned near the ME across a range of disorder strengths.}
        }
	\label{fig4}
\end{figure}

Then, we investigate the $\kappa=3$ photonic mosaic lattice comprising 34 sites at the same modulation strength. The lattice contains four MEs given by the expression of $E_c=\pm J \sqrt{1\pm J/\lambda}$, see~\cite{Wang2020} and SI~A1. The corresponding energy diagram is depicted in Fig.~\ref{fig3}(a), where the four MEs have divided the spectrum into various regions. The inset shows the an enlarged region with 4 MEs at strong modulation strength. Here, we mark three distinct localized eigenstates as \textbf{I, II}, and \textbf{III} (see SI~G for intensity distributions of $\kappa=3$ eigenstates), and we probe the localized states with single site excitation. After a propagation distance of 3000 \textmu m, the evolution pattern distributions are illustrated in Fig.~\ref{fig3}(b) and (c), corresponding to localized states and extended states respectively. The experimental data of localized states demonstrates remarkable agreement with the theoretical simulation results (shown as blank bars in the figure), confirming the exponential localization in the excited site. Moreover, we also inject light into the 16th, 19th and 25th waveguide, to probe the extended state. These particular injection positions are chosen since the site has a large overlap with the two extended eigenstates, survived at such larger potential (also shown in SI~G). Both linear and logarithmic scale intensity distributions are exhibited, and clearly depict a marked contrast to the localized states in terms of light evolution patterns. Our results prove the coexistence of both extended and localized states in the system at different energies, induced by multiple MEs in the $\kappa=3$ mosaic model.

We further confirm the existence of the ME by scanning the quasiperiodic disorder strength in a $\kappa=2$ mosaic lattice and investigating the critical state near ME. First, we would like to point out again that in the photonic platform, it is commonly believed that it is extremely challenging to excite the photonic lattice with eigenstates. Here in our experiment, we chose to use single site excitation to probe the light dynamics. As shown in Fig.~\ref{fig4}(a), we have shown that there are always eigenstates close to the MEs, specifically the 14th and 8th eigenstates (critical ones), while we scan the disorder strength in the range of $\lambda/J=1.5$ to 5. For instance, the 14th eigenstate is a critical state, so the wave packet from single-site excitation at port 21, having the maximal overlap with it, will not be fully localized under its propagation. This suggests that by choosing a small propagation distance where the light is primarily localized at the input port, we can reconstruct the weights even without phase information, due to the dominant light intensity at the input port. We would also like to point out if $\lambda/J$ is small, the photonic lattices actually have very little on-site tuning strength, thus the lattices are very close to a uniform lattice and such measurements are under the continuous-time quantum walks framework, which have been intensively explored (for instance, see Ref.\cite{Perets2008} and Ref.\cite{Tang2018}). The intensity distribution of these eigenstates with $\lambda/J$ = 5 is demonstrated in Fig.~\ref{fig4}(b). To approximate these intensity distributions, we choose port 21 (for the 14th eigenstate) and 17 (for the 8th eigenstate) as our input and probe the light evolution dynamics under a small propagation distance. From the experimental output intensity distribution, we calculated the overlap between our experimental data and different eigenstates. In Fig.~\ref{fig4}(c), it exhibits that our light dynamics always has a high overlap weight with the 14th eigenstates and reasonable overlap with 8th one as we scan the disorder strength. Both panels in Fig.4(c) show good overlaps with the critical states in a whole range of disorder strength (sometimes even better than the bare single-site overlap), which, in turn, allows us to probe the ME.

\section{Conclusion}

In conclusion, we have experimentally implemented a novel class of quasiperiodic mosaic lattices, marking a significant advancement in the quest to understand ME physics. By leveraging integrated photonics platforms, we designed and realized these mosaic lattices in a scalable and flexible manner, enabling rapid prototyping. This approach allowed us to effectively probe the intricate behavior arising from the coexistence of both extended and localized states within the mosaic model, providing valuable insights into the energy-dependent localization transition. Our work demonstrates that quasiperiodic mosaic systems indeed exhibit richer physics than commonly appreciated random disorder models, substantially extending our understanding of the mechanisms driving phase transitions in disordered systems. Our experimental implementation by photonic lattice platform offers a practical means of controlling lattice parameters (such as on-site potential, hopping term~\cite{Andrea2016} and non-Hermiticity~\cite{Song2019}) within a wide tunable range at room temperature to study ME physics~\cite{Zhou2023,Longhi2024}, in contrast to cold atomic systems. Furthermore, preparing samples with different evolution lengths allows observation of wavepacket dynamics, which has been experimentally demonstrated to directly detect topological invariants~\cite{WangSSH,JiaoSSH}. Our findings, therefore, underscore a significant advancement in the field, with the potential to catalyze new research directions in quantum physics, materials science, and beyond.

\section*{Conflict of interest}

The authors declare that they have no conflict of interest.

\section*{Acknowledgements}

The authors would like to thank Prof. David B. Haviland for helpful discussions and suggestions on atomic force microscopy. J.G. acknowledges support from Swedish Research Council (Ref: 2023-06671 and 2023-05288), Vinnova project (Ref: 2024-00466) and the Göran Gustafsson Foundation. A.W.E acknowledges support Knut and Alice Wallenberg (KAW) Foundation through the Wallenberg Centre for Quantum Technology (WACQT), Swedish Research Council (VR) Starting Grant (Ref: 2016-03905), and Vinnova quantum kick-start project 2021. V.Z. acknowledges support from the KAW and VR. I.M.K. acknowledges support from the Russian Science Foundation (grant 21-12-00409). Work at Nordita was supported by  European Research Council under the European Union Seventh Framework ERS-2018-SYG HERO, KAW 2019.0068 and the University of Connecticut.

\section*{Author contributions}

Jun Gao, Ivan M. Khaymovich and Ali W. Elshaari conceived the experiment. Jun Gao and Ze-Sheng Xu performed the measurements. Govind Krishna and Adrian Iovan fabricated and characterized the samples. Jiayidaer Jieensi and Andrea Cataldo analyzed the AFM data. Jun Gao, Xiao-Wei Wang, Ivan M. Khaymovich and Ali W. Elshaari discussed and analyzed the results. Jun Gao made the figures with inputs from Ivan M. Khaymovich, Xiao-Wei Wang and Ali W. Elshaari. Jun Gao, Ali W. Elshaari and Ivan M. Khaymovich wrote the manuscript. Val Zwiller, Alexander V. Balatsky and Ali W. Elshaari supervised the project.

\end{document}